\documentclass[sn-nature]{sn-jnl}

\usepackage{graphicx}%
\usepackage{amsmath,amssymb,amsfonts}%
\usepackage{textcomp}%
\usepackage{manyfoot}%
\usepackage{xcolor}%
\usepackage{setspace}%
\usepackage{lineno}%
\usepackage{bm}

\newcommand{\NiFe}{Ni\textsubscript{\emph{x}}Fe\textsubscript{100\(-\)\emph{x}}}

\makeatletter
\def\abstractfont{\reset@font\fontsize{10bp}{12bp}\selectfont\unboldmath\leftskip=24pt\rightskip=24pt\parfillskip=0pt plus 1fil}%
\def\keywordfont{\reset@font\fontsize{10bp}{12bp}\selectfont\leftskip=24pt\rightskip=24pt plus0.5fill}%
\makeatother

\begin{document}

\title{\vspace{-1\baselineskip}Nonreciprocity reversal of magnetoacoustic attenuation in NiFe alloy thin films}

\author*[1,2,3,7]{\fnm{Mingran} \sur{Xu}}\email{xu.mingran.d8@tohoku.ac.jp}
\equalcont{These authors contributed equally to this work.}

\author[4,3]{\fnm{Kei} \sur{Yamamoto}}
\equalcont{These authors contributed equally to this work.}

\author*[3,5]{\fnm{Kouta} \sur{Kondou}}\email{kondou.kouta.otri@osaka-u.ac.jp}

\author[1]{\fnm{Zheng} \sur{Zhu}}

\author[1]{\fnm{Liyang} \sur{Liao}}

\author[3]{\fnm{Kiyohiro} \sur{Adachi}}

\author[3]{\fnm{Tomoka} \sur{Kikitsu}}

\author[3]{\fnm{Daisuke} \sur{Hashizume}}

\author[2,6]{\fnm{Dirk} \sur{Grundler}}

\author[3,4]{\fnm{Sadamichi} \sur{Maekawa}}

\author*[1,3,7]{\fnm{Yoshichika} \sur{Otani}}\email{yoshichika.otani.b7@tohoku.ac.jp}

\affil[1]{\orgdiv{Institute for Solid State Physics}, \orgname{University of Tokyo}, \orgaddress{\street{5-1-5 Kashiwanoha}, \city{Kashiwa}, \state{Chiba}, \postcode{277-8581}, \country{Japan}}}

\affil[2]{\orgdiv{Laboratory of Nanoscale Magnetic Materials and Magnonics, Institute of Materials}, \orgname{\'Ecole Polytechnique F\'ed\'erale de Lausanne}, \orgaddress{\city{Lausanne}, \postcode{1015}, \country{Switzerland}}}

\affil[3]{\orgdiv{Center for Emergent Matter Science (CEMS)}, \orgname{RIKEN}, \orgaddress{\city{Saitama}, \postcode{351-0198}, \country{Japan}}}

\affil[4]{\orgdiv{Advanced Science Research Center}, \orgname{Japan Atomic Energy Agency}, \orgaddress{\city{Tokai}, \postcode{319-1195}, \country{Japan}}}

\affil[5]{\orgdiv{Institute for Open and Transdisciplinary Research Initiatives}, \orgname{The University of Osaka}, \orgaddress{\city{Osaka}, \postcode{565-0871}, \country{Japan}}}

\affil[6]{\orgdiv{Institute of Electrical and Micro Engineering}, \orgname{\'Ecole Polytechnique F\'ed\'erale de Lausanne}, \orgaddress{\city{Lausanne}, \postcode{1015}, \country{Switzerland}}}

\affil[7]{\orgdiv{Department of Physics}, \orgname{Tohoku University}, \orgaddress{\city{Sendai}, \postcode{980-8578}, \country{Japan}}}

\abstract{Nonreciprocity, the asymmetry of transport, underlies technologies from the diode to the microwave isolator. In a ferromagnet, a surface acoustic wave generates an elliptical effective field with propagation-locked handedness, breaking the reciprocity of its propagation. Despite decades of study on this phenomenon, a method for controlling the sign of the nonreciprocity has remained elusive. Here we observe a sign reversal in \NiFe{} films. A \(0.8\)~at.\% change across Permalloy's zero-magnetostriction composition, where the magnetoelastic coefficient \(b\) changes sign, reverses the handedness of the elliptical effective field and thereby the nonreciprocity, from \(78.6\%\) to \(-61.8\%\). Angle-dependent measurements and spin-wave-ellipticity modelling show that reversing the sign of \(b\) reverses the handedness of the elliptically polarized effective field. Aided by cubic frequency scaling, we resolve the sign of \(b\) down to \(-0.05\)~MPa in a \(10\)-nm film, establishing nonreciprocity as a nanoscale probe of magnetoelastic coupling.}
\keywords{Magnon--phonon coupling, Surface acoustic waves, Nonreciprocity, Magnetoelastic coupling, Permalloy alloys}

\maketitle

\hspace{15pt}Magnons and phonons are promising carriers for wave-based logic, processing information as waves rather than charge and so avoiding Joule heating~\cite{Chumak2015}. Magnon--phonon coupling~\cite{Matthews1962}, which links magnetization dynamics to lattice vibrations, is of broad interest across spintronics~\cite{Wieler2,Kikkawa2016,KawadaHayashi}, magnonics~\cite{Jamison,JileiSAW,kyongmoAn}, acoustics~\cite{chenchong}, and quantum technology~\cite{hwang}, uniting the complementary strengths of the two subsystems~\cite{Ganguly1976,kunstel2025}: phonons propagate over long distances for on-chip signal transport, including coherent quantum-information transfer~\cite{Bienfait2019}, while magnons write and read spin information through magnetic elements~\cite{Baumgaertl2023,Yokouchi2020}.

In magnon--phonon coupling driven by Rayleigh-type surface acoustic waves (SAWs)~\cite{Wieler,JileiSAW,kunstel2025}, a pronounced nonreciprocity between forward- and backward-propagating SAWs (SAW\(_{\text{+}k}\) and SAW\(_{\text{-}k}\)) was observed in the 1970s~\cite{Daniel1977a} and still attracts strong interest~\cite{Onose,Matthias2021}. Such nonreciprocal transmission could integrate the microwave isolator function~\cite{Lewis1972} into SAW filters, essential components of microwave technology~\cite{Chen2024SAF}; but a wave-based logic platform also needs signals actively routed between channels by a circulator~\cite{Shao2020}, whose operation relies on a definite sign of the nonreciprocity, a materials-level lever that has so far remained elusive.

The origin of the nonreciprocity has been assigned to an elliptically polarized effective field, driven by the SAW, with propagation-locked handedness~\cite{Xu2020,Yamamoto2020,LopesSeeger2024}. The in-plane component of this field is driven by the longitudinal strain of the SAW through the magnetoelastic (ME) coupling with coefficient \(b\), while the out-of-plane component receives two contributions, one from the shear strain, again through the ME coupling, and one from the lattice rotation through the magneto-rotation (MR) coupling~\cite{Maekawa1976,Xu2020,Yamamoto2020}, whose strength is set by the effective magnetic anisotropy \(K\). The two components oscillate with a \(\pm\pi/2\) phase offset, and their relative sign, fixed by the propagation direction, sets the handedness of the drive; because the magnetization itself precesses with one fixed handedness, SAW\(_{\text{+}k}\) and SAW\(_{\text{-}k}\) are absorbed with different strengths. Crucially, we expect that the two out-of-plane drives respond differently to a sign change of \(b\). In an isotropic polycrystalline magnet, whose randomly oriented crystallites average to a single magnetoelastic coefficient, the shear strain shares that coefficient \(b\) with the longitudinal strain~\cite{duTremolet1993}, leaving the drive's handedness untouched, whereas the lattice rotation does not, inverting it.

Here, we demonstrate composition-controlled, systematic changes in the sign and magnitude of magnetoacoustic nonreciprocity in a series of \NiFe{} thin films integrated onto nominally identical SAW delay lines (Fig.~\ref{fig1}a). A mere \(0.8\)~at.\% change in Ni content \emph{x} near the zero-magnetostriction composition of Permalloy (\emph{x}~=~80)~\cite{McKeehanSig1926,Chikazumi1997,MEBoin}, where \(b\) changes sign, reverses the nonreciprocity from \(78.6\%\) to \(-61.8\%\) (Fig.~\ref{fig2}). Microscopically, this sign reversal originates from the change in the strain dependence of the spin--orbit-induced magnetic anisotropy~\cite{Chikazumi1997,Daalderop1990}. Around the zero-magnetostriction composition of \NiFe{}, competing orbital contributions from Ni and Fe nearly cancel, causing the magnetostriction coefficient, and therefore the magnetoelastic coefficient \(b\), to change sign. A theoretical model quantitatively reproduces the field-angle dependence of the attenuation (Fig.~\ref{fig3}), with the extracted degree of nonreciprocity tracking the same sign change of \(b\). In the frequency-dependent measurements, a refined treatment of the spin-wave ellipticity shows that the two rotation-based mechanisms, the MR coupling and the Barnett field mechanism, both reproduce the measured cubic frequency scaling of the attenuation (Fig.~\ref{fig4}).
The observed reversal identifies a rotation-based, rather than shear-strain-based, origin of the nonreciprocity. Because this contribution does not scale with \(b\), the response persists as \(b\to 0\). Indeed, at elevated SAW frequencies, near-zero-magnetostriction Permalloy exhibits an attenuation one order of magnitude larger than previously reported~\cite{Kurimune2020}, establishing rotation-mediated coupling as an efficient handle on weakly magnetoelastic systems.

The robustness of the nonreciprocity against vanishing \(b\) turns it into a nanoscale, sign-sensitive probe of the magnetoelastic coefficient. Read out electrically from the \(10\)-nm-thick device film, which contains four orders of magnitude less magnetic volume than bending-based metrology requires~\cite{Klokholm1976,MEBoin,Greenall2024}, it resolves the sign of \(b\) down to \(b \approx -0.05\)~MPa (saturation magnetostriction \(\lambda_{s} \approx +0.2\)~ppm) in Ni\textsubscript{79.9}Fe\textsubscript{20.1}, a regime where bending-based metrology loses signal. This is precisely the weakly magnetoelastic regime engineered for magnetoresistive random-access memory (MRAM)~\cite{Engel2005,Sankaran2018} and sensors~\cite{KlokholmAboaf1981,Freitas2007}, where the sign of the residual \(b\) sets the direction of the compositional correction. The same compositional control over sign and magnitude also broadens the design space for nonreciprocal magnon--phonon devices such as isolators and circulators with reconfigurable strength and directionality~\cite{Shao2020}.

\section*{Observation of nonreciprocity reversal}

\begin{figure*}[htp]
\centering
\includegraphics[width=\textwidth]{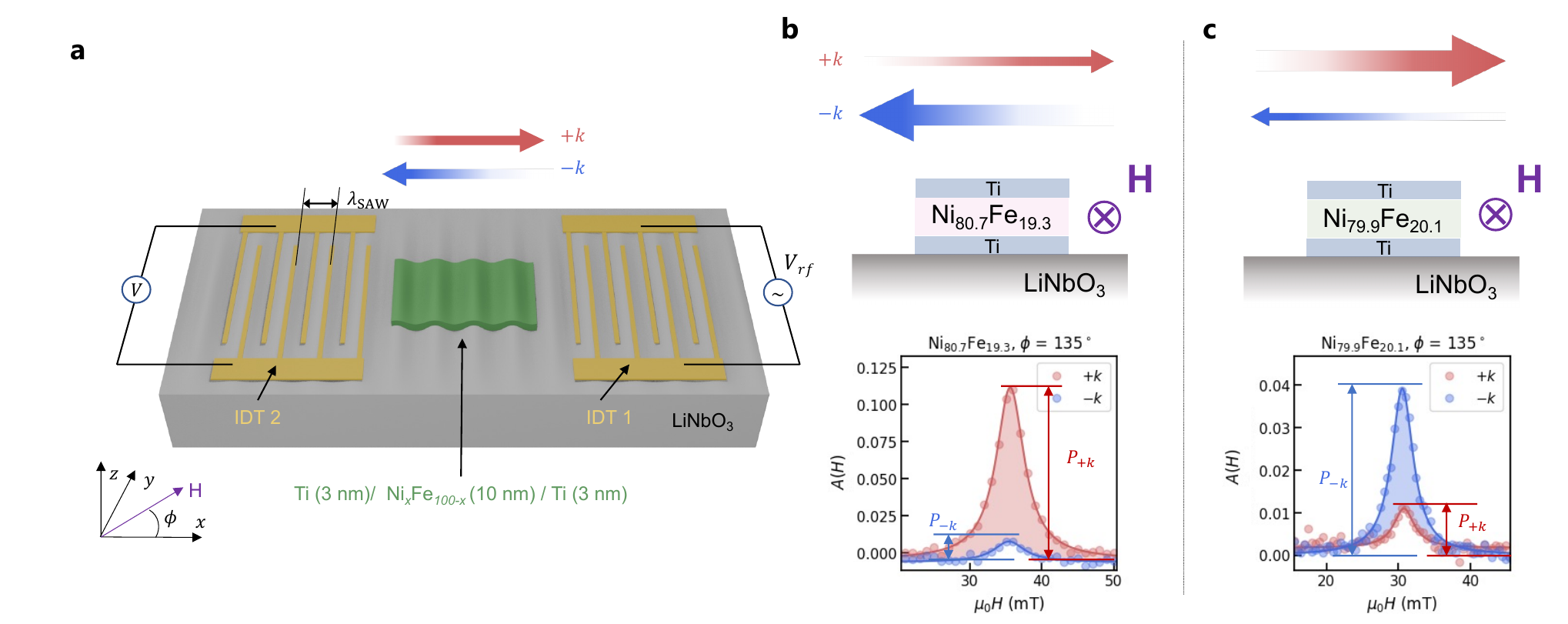}
\caption{\textbf{Nonreciprocity reversal of magnetoacoustic attenuation in a hybrid device.} a, Schematic of the hybrid device: two interdigital transducers (IDT 1 and IDT 2) on a LiNbO\textsubscript{3} substrate form a SAW delay line, with a Ti (3 nm)/\NiFe{} (10 nm)/Ti (3 nm) thin film patterned on the acoustic propagation path between the IDTs. SAWs launched from either IDT propagate in the \(+k\) or \(-k\) direction and couple to the film. An in-plane magnetic field \emph{\textbf{H}} is applied at an angle \emph{\(\phi\)} from the SAW propagation direction (the \emph{x} axis). b, c, Upper schematics summarize the propagation-direction-dependent magnetoacoustic attenuation under \emph{\textbf{H}}, illustrating which SAW direction (\(+\)\emph{k} or \(-\)\emph{k}) is more strongly attenuated for each composition. Lower panels show magnetoacoustic attenuation spectra \emph{A(H)} at \emph{\(\phi\)} = 135\textdegree{} for Ni\textsubscript{80.7}Fe\textsubscript{19.3} (b) and Ni\textsubscript{79.9}Fe\textsubscript{20.1} (c), comparing SAW\(_{\text{+}k}\) and SAW\(_{\text{-}k}\); the nonreciprocal contrast reverses between the two compositions. Shaded regions highlight the difference between the \(+\)\emph{k} and \(-\)\emph{k} responses. The extracted peak absorptions are \(P_{\text{+}k} = 0.1165\), \(P_{\text{-}k} = 0.0140\) for Ni\textsubscript{80.7}Fe\textsubscript{19.3} and \(P_{\text{+}k} = 0.0093\), \(P_{\text{-}k} = 0.0393\) for Ni\textsubscript{79.9}Fe\textsubscript{20.1}, giving \(\chi_{P} = 78.6\%\) and \(-61.8\%\). The resonance fields differ between \textbf{b} and \textbf{c} because the saturation magnetization varies with composition (Supplementary Fig.~S2).}
\label{fig1}
\end{figure*}

The hybrid device consists of two interdigital transducers (IDTs) forming a SAW delay line on 128\textdegree{} Y-X cut LiNbO\textsubscript{3}, with a Ti/\NiFe{}/Ti trilayer patterned on the acoustic path (Fig.~\ref{fig1}a; see Methods). The symmetric top and bottom interfaces suppress the interfacial Dzyaloshinskii--Moriya interaction and thereby its contribution to the nonreciprocity~\cite{Kuss2020,Xu2020}. The IDT periodicity of 630~nm sets the acoustic resonance frequency \(f_{0} = 6.2\)~GHz.

An in-plane magnetic field \(\bm{H}\) of magnitude \(H\) and in-plane angle \(\phi\) (Fig.~\ref{fig1}a, which defines the axes) brings the spin-wave mode of wavelength \(\lambda_{\text{SAW}}\) into resonance with the SAW, transferring energy from the acoustic to the magnetic excitation. The absorption is quantified by the magnetoacoustic attenuation \(A_{\pm k}(H,\phi)\), the fractional reduction of transmitted acoustic power, obtained from raw transmission signals for SAW\(_{\pm k}\) (see Methods). The attenuation (Fig.~\ref{fig1}b, c) is well characterized by

\begin{equation}\label{eq:main1}
A_{\pm k}(H,\phi) = \frac{\Delta(\phi)^{2}\,P_{\pm k}(\phi)}{\left( H - H_{\text{res}}(\phi) \right)^{2} + \Delta(\phi)^{2}} + R_{\pm k}(\phi),
\end{equation}
where \(H_{\text{res}}\) and \(\Delta\) are the resonance field and half-width at half-maximum of the spin-wave mode, \(P_{\pm k}\) is the peak absorption amplitude above the background, and \(R_{\pm k}\) is the $H$-independent background.

Representative data from two devices (Ni\textsubscript{80.7}Fe\textsubscript{19.3} and Ni\textsubscript{79.9}Fe\textsubscript{20.1}), measured by sweeping \(H\) down from 70 mT at \(\phi = 135\textdegree{}\) (near the angle of maximal magnetoelastic drive), already reveal the effect: differing by only 0.8~at.\% in Ni content, they show a clear reversal of the nonreciprocity, with SAW\(_{\text{+}k}\) more strongly attenuated in Ni\textsubscript{80.7}Fe\textsubscript{19.3} and SAW\(_{\text{-}k}\) in Ni\textsubscript{79.9}Fe\textsubscript{20.1} (Fig.~\ref{fig1}b, c).

\section*{Nonreciprocity across the zero-magnetostriction composition}

\begin{figure*}[htp]
\centering
\includegraphics[width=\textwidth]{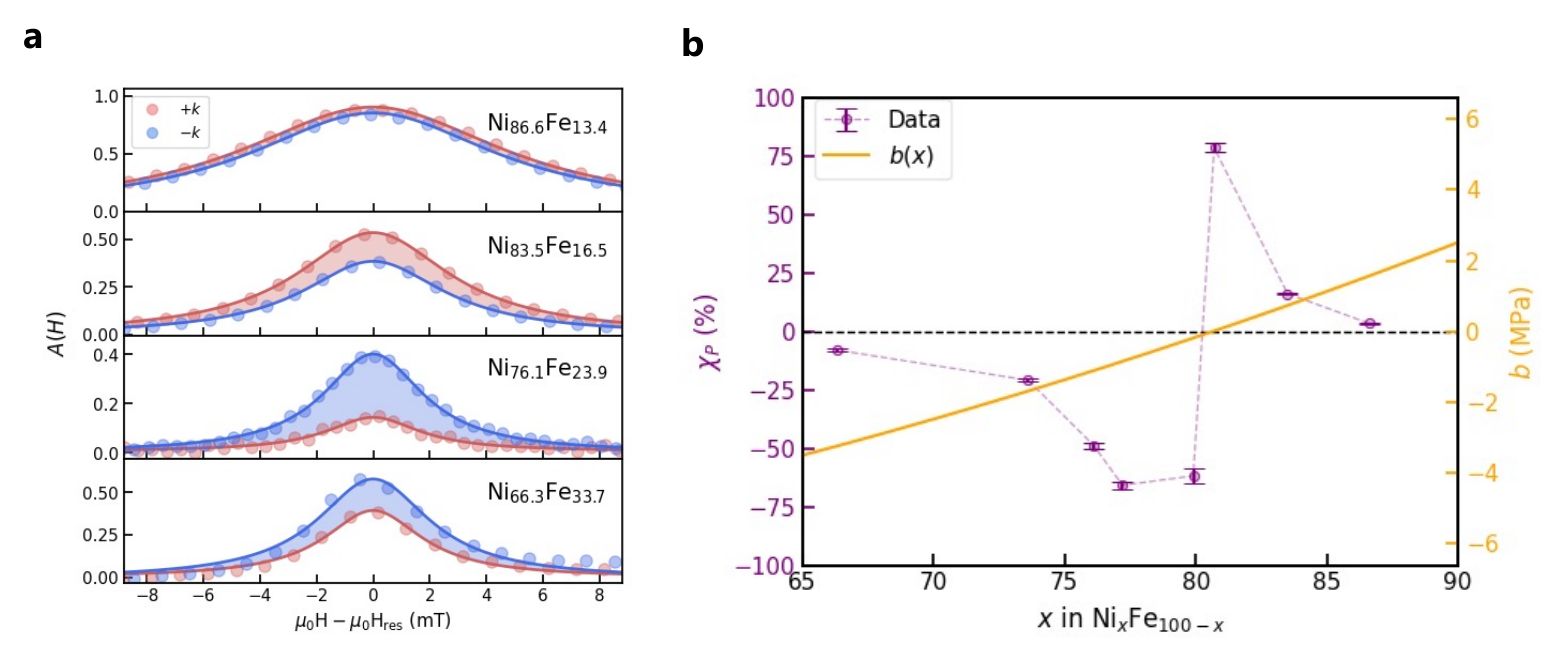}
\caption{\textbf{Composition-driven reversal of magnetoacoustic attenuation nonreciprocity in NiFe alloys.} \textbf{a,} Magnetoacoustic attenuation spectra \emph{A(H)} measured at room temperature near the spin-wave resonance for counter-propagating SAW\(_{\pm k}\) in \NiFe{} at \(\phi = 135^{\circ}\). Shaded regions highlight the difference between the \(+k\) and \(-k\) responses. \textbf{b,} Nonreciprocity ratio \(\chi_{P}\) as a function of Ni concentration, shown together with the composition-dependent coefficient \emph{b} (calculated from literature values~\cite{MEBoin,Shirakawa1969}; details are given in the Supplementary Information). \(\chi_{P}\) reverses sign near \emph{x} = 80 in \NiFe{}, coincident with the sign change of \emph{b}. Error bars represent fitting uncertainties. The closed-form expression \emph{b}(\emph{x}) = 1.806 \(\times\) 10\textsuperscript{\(-\)3} \emph{x}\textsuperscript{2} \(-\) 3.956 \(\times\) 10\textsuperscript{\(-\)2} \emph{x} \(-\) 8.554 MPa is derived in Supplementary Eq. (S61).}
\label{fig2}
\end{figure*}

We studied a series of such devices of varying \emph{x} at a nominally identical SAW wavenumber (\(\sim\)1.0 \(\times\) 10\textsuperscript{7} rad/m), sweeping \(H\) at \(\phi\) = 135\textdegree{} for both SAW\(_{\text{+}k}\) (red dots) and SAW\(_{\text{-}k}\) (blue dots), with the horizontal axis referenced to \(\mu_{0}H_{\text{res}}\) for each device (Fig.~\ref{fig2}).

We define the nonreciprocity ratio \(\chi_{P} = \left( P_{\text{+}k} - P_{\text{-}k} \right)/\left( P_{\text{-}k} + P_{\text{+}k} \right)\) (Fig.~\ref{fig2}b).
It is non-monotonic in \emph{x}, peaking in magnitude near \emph{x}~=~80 and reversing sign there, coincident with the sign change of \emph{b} (yellow curve). Voigt--Reuss--Hill averaging of the single-crystal elastic constants~\cite{Shirakawa1969} together with the composition-dependent magnetostriction~\cite{MEBoin} yields the \emph{b}(\emph{x}) plotted in Fig.~\ref{fig2}b (Supplementary Eq.~(S61)); built from independently published data, this model is intended to capture the trend in \emph{b}(\emph{x}), chiefly its zero crossing and local slope, rather than to serve as a quantitative benchmark.

This correlation follows from the deformations carried by a Rayleigh-type SAW: a longitudinal strain \(\varepsilon_{xx} = \partial u_{x}/\partial x\), a shear strain \(\varepsilon_{xz} = \tfrac{1}{2}\left(\partial u_{x}/\partial z + \partial u_{z}/\partial x\right)\), and a lattice rotation \(\omega_{xz} = \tfrac{1}{2}\left(\partial u_{x}/\partial z - \partial u_{z}/\partial x\right)\), where \(u_{i}\) (\(i = x, y, z\)) are the Cartesian components of the elastic displacement field (Fig.~\ref{fig1}a). These deformations drive the effective field with in-plane and out-of-plane components \(h_{\mathrm{IP}} \propto b\,\varepsilon_{xx}\) and \(h_{\mathrm{OOP}} \propto b\,\varepsilon_{xz},\ K\omega_{xz}\), which oscillate with a \(\pm\pi/2\) phase difference, so the handedness of the drive is set by the triad of the SAW's propagation direction, the film normal, and \(\bm{H}\). For the shear-strain mechanism, the isotropic magnetoelastic response of the polycrystal gives \(h_{\mathrm{IP}}\) and \(h_{\mathrm{OOP}}\) the same coefficient \(b\), so a sign change of \(b\) flips both components together, a global \(\pi\) phase shift of the drive that leaves the drive's handedness unchanged. For the MR mechanism, by contrast, we assume the anisotropy is dominated by shape anisotropy, \(K = -\mu_{0}M_{\mathrm{S}}^{2}/2\) with \(M_{\mathrm{S}}\) the saturation magnetization, so that \(K\) is independent of \(b\); a sign change of \(b\) then reverses the sign of \(h_{\mathrm{OOP}}/h_{\mathrm{IP}}\) and thereby its handedness. Continuum elasticity further predicts the shear drive to be small compared with the rotation, since the stress-free surface suppresses \(\varepsilon_{xz}\) relative to \(\omega_{xz}\) by \(\sim kd\) for a nanometre-thick surface film (see Supplementary Information). We do not rely on this expectation, since the sign-reversal argument holds whatever the shear magnitude, but, as will be seen, the consistency of the rotation-based analysis across all compositions bears it out. The observed reversal of \(\chi_{P}\) at the sign change of \(b\) is therefore the fingerprint of the rotation-based mechanism, which we quantify in the next section.

\section*{Model of nonreciprocal magnetoacoustic attenuation}

Several mechanisms can produce nonreciprocal attenuation~\cite{Xu2020}, but we focus on the shear-strain contribution of the ME coupling and the MR coupling in this section. Refining that treatment to give the frequency-dependent spin-wave ellipticity its full weight (see Supplementary Information), we find that both yield the same functional form:

\begin{equation}\label{eq:main2}
P_{\pm k} = \frac{C_{0}f^{2}}{\kappa_{\text{SW}}}\frac{m_{\text{ip}}}{m_{\text{oop}}}\left( \frac{1}{\delta} \mp \sin\phi \right)^{2}\cos^{2}\phi,\qquad C_{0}\propto \frac{b^{2}}{M_{\text{S}}}.
\end{equation}
Here \(f\) is the SAW frequency, \(\kappa_{\text{SW}}\) the spin-wave relaxation rate, and \(m_{\text{ip}}\), \(m_{\text{oop}}\) the in- and out-of-plane precession amplitudes. The prefactor \(C_{0}\) is independent of \(H\), \(\phi\), \(\lambda_{\text{SAW}}\), and \(f\), its magnetic-material dependence entering through the \(b^{2}/M_{\text{S}}\) scaling. Within the bracket, the \(\sin\phi\) term is the in-plane longitudinal-strain drive and
the dimensionless nonreciprocity parameter \(\delta\) is the ratio of the in-plane to the out-of-plane drive; interfering with opposite sign for \(+k\) and \(-k\) (the \(\mp\)), they give \(\chi_{P} = -2\delta\sin\phi/(1 + \delta^{2}\sin^{2}\phi)\). A finite \(\delta\) renders the response nonreciprocal, and the sign of \(\delta\) sets that of \(\chi_{P}\).

The two mechanisms differ in what determines \(\delta\), as derived in the Supplementary Information:
\begin{equation}\label{eq:main3}
\delta = \begin{cases}
\ 2\dfrac{L_{\bm{k}}}{S_{\bm{k}}}\,\dfrac{m_{\text{ip}}}{m_{\text{oop}}}, & \text{shear-strain mechanism} \\[10pt]
-\dfrac{4b\eta}{\mu_{0}M_{\text{S}}^{2}}\,\dfrac{m_{\text{ip}}}{m_{\text{oop}}}, & \text{MR mechanism}
\end{cases}
\end{equation}
where \(\eta = L_{\bm{k}}/W_{\bm{k}}\) is an order-unity constant set by the SAW mode profile, with \(L_{\bm{k}}\), \(S_{\bm{k}}\), and \(W_{\bm{k}}\) dimensionless measures of the longitudinal strain, shear strain, and lattice rotation of the Rayleigh mode, averaged over the film thickness (see Supplementary Information); since \( kd \ll 1\) in our devices, these deformations are approximately uniform through the film thickness,
so the film-averaged and surface values coincide and the constants reduce to \(L_{\bm{k}}/S_{\bm{k}} \approx \varepsilon_{xx}/\varepsilon_{xz}\) and \(\eta \approx \varepsilon_{xx}/\omega_{xz}\), where $\varepsilon _{xx}, \varepsilon _{xz} , \omega _{xz}$ are evaluated at the surface. The shear-strain \(\delta\) carries no magnetoelastic coefficient, since for the isotropic polycrystalline films the common \(b\) cancels between the shear and longitudinal drives; its sign is therefore independent of the NiFe composition, and \(\chi_{P}\) is not expected to reverse. The MR term, by contrast, carries \(b\) explicitly, so the sign of \(\delta\), and hence of the nonreciprocity, follows the sign of \(b\); the observed reversal at the zero-magnetostriction composition is consistent with this picture, pointing to a rotation-based, rather than shear-strain-mediated, origin of the nonreciprocity.

For an in-plane magnetized thin film the precession is strongly oblate~\cite{GurevichMelkov1996},
\begin{equation}\label{eq:main4}
\frac{m_{\text{ip}}}{m_{\text{oop}}} \approx \frac{|\gamma|\mu_{0}M_{\text{S}}}{2\pi f} \gg 1,
\end{equation}
with \(\gamma\) the gyromagnetic ratio, because the demagnetizing field stiffens the out-of-plane precession relative to the in-plane one (see Supplementary Information). This ellipticity re-weights the in-plane and out-of-plane drives and governs the frequency dependence of the attenuation, examined in the sections that follow.

\section*{Field-angle analysis as a probe of the sign of \(b\)}

\hspace{15pt}Equation~(2) predicts an attenuation that varies with in-plane field angle as $\left( 1/\delta \mp \sin\phi \right)^{2}\cos^{2}\phi$. We fit this profile to the measured angular-dependent attenuation to determine $\delta$ for each composition.

\begin{figure*}[htp]
\centering
\includegraphics[width=\textwidth]{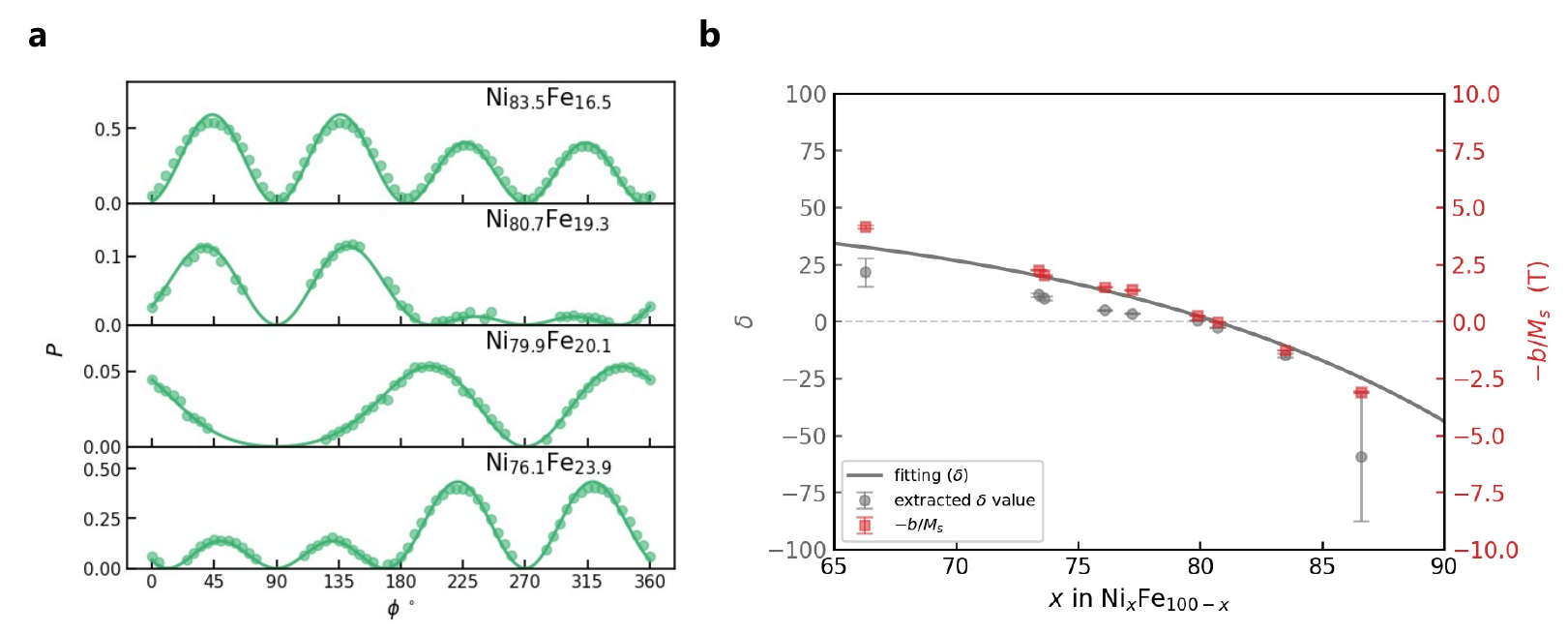}
\caption{\textbf{Magnetic-field-angle dependence of magnetoacoustic attenuation in \NiFe{} alloys.}
\textbf{a,} Angular dependence of the magnetoacoustic attenuation peak amplitude \(P\) for \NiFe{} alloys of varying composition; \(\phi = 90\textdegree{}\) corresponds to the Damon--Eshbach configuration.
\textbf{b,} Composition dependence of the fitted parameter \(\delta\) (grey circles, left axis), extracted from the angular analysis in \textbf{a}; error bars denote fitting uncertainties. The solid grey line is a fit to \(\delta(x) = \beta\,b(x)/M_\mathrm{S}(x)\), with \(\beta\) as the only free parameter. Red squares (right axis) show \(-b/M_\mathrm{S}\).}
\label{fig3}
\end{figure*}

We extracted the peak absorption \(P(\phi)\) from Lorentzian fits based on Eq.~(\ref{eq:main1}) over the full in-plane angular range (Fig.~\ref{fig3}a; see Methods).  

As the composition varies, \(P(\phi)\) is well tracked by a gradual change of \(\delta\) in Eq.~(\ref{eq:main2}): from an approximately fourfold \(\sin^{2}\phi\cos^{2}\phi\) dependence for the most Ni-rich sample (Ni\textsubscript{83.5}Fe\textsubscript{16.5}), through a pronounced mixed \(\sin\phi\cos^{2}\phi\) term near Ni\textsubscript{80.7}Fe\textsubscript{19.3}, to a \(\cos^{2}\phi\)-dominated form at Ni\textsubscript{79.9}Fe\textsubscript{20.1} where \(b \to 0\), and finally to a sign-reversed \(-\sin\phi\cos^{2}\phi\) distortion at Ni\textsubscript{76.1}Fe\textsubscript{23.9}. The extracted \(\delta\) thus varies systematically with \emph{x} and inherits the sign change of \emph{b} near \emph{x}~=~80 (Fig.~\ref{fig3}b).  
 
Combining the MR \(\delta\) of Eq.~(\ref{eq:main3}) with the ellipticity of Eq.~(\ref{eq:main4}) shows that, at fixed frequency, \(\delta \propto -b/M_{\mathrm{S}}\): the factor of \(M_{\mathrm{S}}\) supplied by the ellipticity cancels the \(M_{\mathrm{S}}^{2}\) in the MR denominator. Overlaying \(-b/M_{\mathrm{S}}\) (red squares, right axis of Fig.~\ref{fig3}b) on the extracted \(\delta(x)\) reproduces its composition dependence, including the sign change near \emph{x}~=~80; a single-parameter fit \(\delta(x) = \beta\,b(x)/M_{\mathrm{S}}(x)\), in which \(b(x)\) and \(M_{\mathrm{S}}(x)\) are fixed by the independent data of Supplementary Eqs.~(S61) and (S62) and \(\beta\) is the only free parameter, yields \(\beta = -8.24 \pm 2.20~\mathrm{T}^{-1}\) (fit quality in the Supplementary Information). This \(\beta\) corresponds to \(\eta \approx -\beta\,\pi f/(2|\gamma|) \approx 0.456 \pm 0.122\),
consistent within its uncertainty with the range \(\varepsilon_{xx}/\omega_{xz} \approx 0.33\text{--}0.40\) obtained from a numerical model of the Rayleigh SAW on 128\textdegree{} Y-X cut LiNbO\textsubscript{3}~\cite{Kawada2025SAW}.

Beyond identifying the mechanism, the angular analysis makes the nonreciprocity a probe of the magnetoelastic coupling whose defining strength is determining the sign of \(b\) at values of \(|b|\) too small for its magnitude to be resolved. The sign follows from \(\text{sign}(b) = \text{sign}(\chi_{P}\sin\phi) = -\text{sign}(\delta)\), read directly from which of SAW\(_{\pm k}\) is more strongly attenuated, without any fitting, and independently of \(\eta\), \(M_{\text{S}}\), or which rotation-based mechanism is operative. Because these enter the magnitude \(b \approx -\delta\,\pi f M_{\text{S}}/(2|\gamma|\eta)\) only as positive multiplicative factors, their uncertainties broaden the error bar on \(|b|\) but cannot change its sign. Inverting Eq.~(\ref{eq:main3}) gives an on-chip estimate \(b \approx -0.05\)~MPa (\(\lambda_{s} \approx +0.2\)~ppm) for Ni\textsubscript{79.9}Fe\textsubscript{20.1}, which assumes the MR mechanism (a Barnett origin rescales it by a factor of two, as discussed in the next section) and reflects the film's actual strain state rather than a fixed material constant. What distinguishes the method from dedicated magnetostriction metrology~\cite{Greenall2024,Klokholm1976} is therefore not accuracy but reach: the sign stays resolvable as \(b \to 0\), precisely where the deflection signal of strain-based methods is expected to fall below the noise floor while the nonreciprocity stays large (\(\chi_{P} = -61.8\%\)).

\section*{Frequency scaling and the two rotation-based mechanisms}

\hspace{15pt}While the analysis so far has been consistent with the MR mechanism, it is not the only rotation-based spin--lattice coupling. The Barnett field, arising from the nonzero vorticity of the Rayleigh SAW, i.e., the time derivative of the lattice rotation \(\partial_t\omega_{xz}\), may equally explain the magnetoacoustic attenuation dependences on Ni composition and \(\phi\). In this spin--vorticity picture the Barnett field~\cite{Kurimune2020} contributes a nonreciprocity parameter~\cite{Xu2020}
\begin{equation}\label{eq:main5}
\delta = -\frac{|\gamma|\,b\,\eta}{\pi f M_{\text{S}}}, \qquad \text{Barnett field mechanism}.
\end{equation}
This \(\delta\) is independent of the field angle, so it yields the same \(\phi\)-dependence of the attenuation [Eq.~(\ref{eq:main2})] as the MR coupling; the angular analysis of the previous section is therefore equally consistent with either, leaving the frequency dependence as the only possible discriminant. Following the suggestion~\cite{Kurimune2020a} that the Barnett field depends more strongly on the SAW frequency than the ME coupling, owing to its additional time derivative, we measured a series of devices with IDTs of varying periodicity spanning acoustic resonance frequencies from \(\sim\)2 to \(\sim\)9 GHz (see Methods).

\begin{figure*}[htp]
\centering
\includegraphics[width=\textwidth]{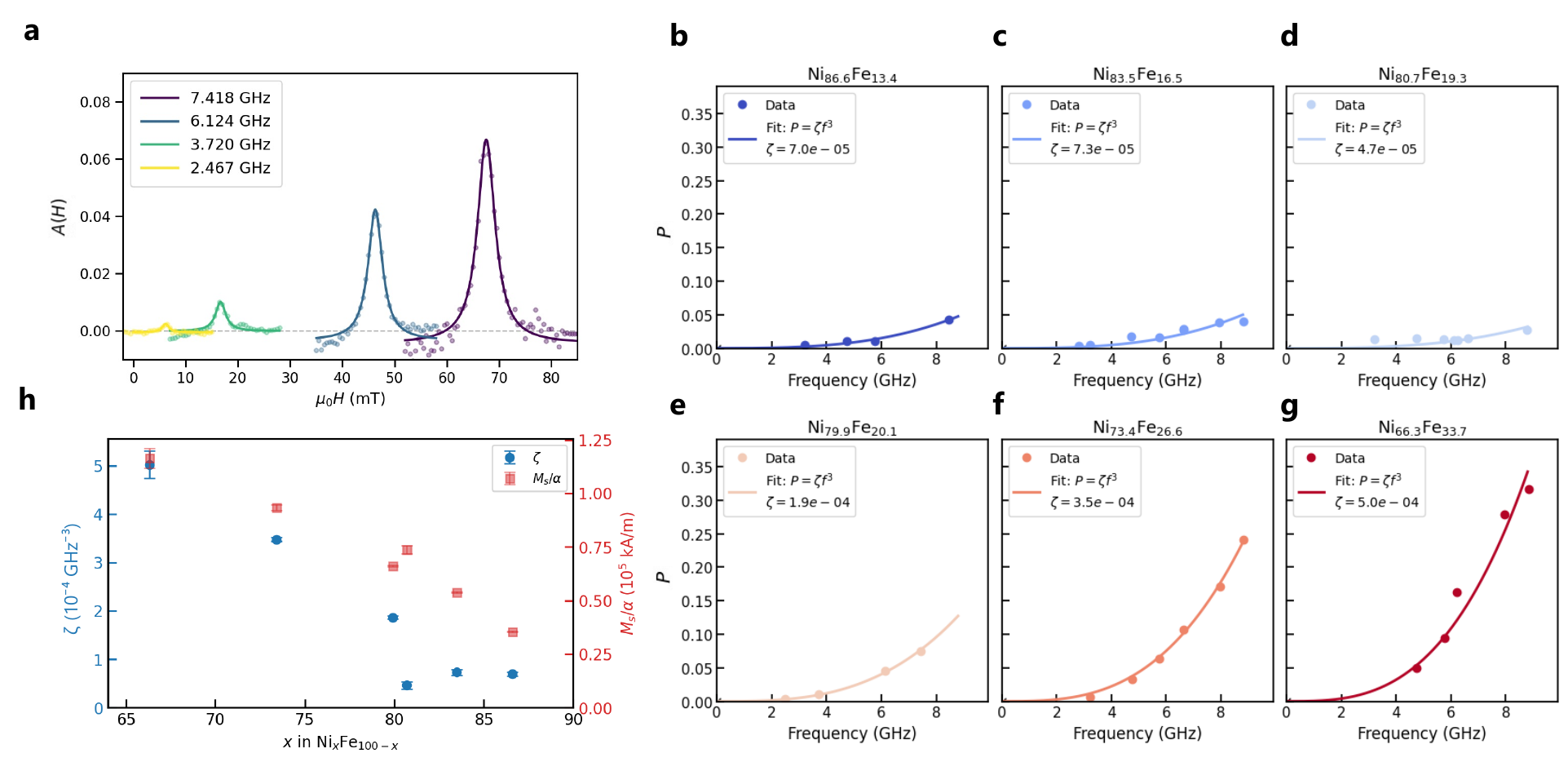}
\caption{\textbf{Frequency dependence of magnetoacoustic attenuation.} \textbf{a,} Representative magnetic-field-dependent magnetoacoustic attenuation spectra acquired at multiple excitation frequencies (indicated in the labels) for the Ni\textsubscript{79.9}Fe\textsubscript{20.1} device at \(\phi = 0^{\circ}\) for SAW\(_{\text{+}k}\); the peak amplitude \(P_{\text{+}k}\) is extracted from Lorentzian fits. \textbf{b--g,} Frequency dependence of \(P_{\text{+}k}\) for the \NiFe{} compositions indicated in each panel (symbols). Solid lines are fits to \emph{P}(\emph{f}) = \(\zeta f^{3}\), and the corresponding fit parameters are shown in the legends; \(\zeta\) is given in GHz\textsuperscript{\(-\)3} throughout. \textbf{h,} Composition dependence of the fitted coefficient \(\zeta\) extracted from panels \textbf{b--g} (error bars denote fitting uncertainties). The composition dependence in \textbf{h} reflects the scaling \emph{\(\zeta\)} \(\propto\) \emph{M}\textsubscript{S}(\emph{x})/\emph{\(\alpha\)}(\emph{x}) given in Supplementary Eq. (S57), where the Gilbert damping \emph{\(\alpha\)}(\emph{x}) is independently extracted from the frequency dependence of the resonance linewidth, as shown in Supplementary Fig.~S3.}
\label{fig4}
\end{figure*}

To isolate the rotation-based mechanisms, spectra were measured at \(\phi = 0^{\circ}\), where the longitudinal-strain (ME) drive, proportional to \(\sin\phi\), vanishes; the remaining attenuation is dominated by the rotation-based coupling (the shear drive being small, as discussed above) and, to that extent, independent of \(b\). For a Ni\textsubscript{79.9}Fe\textsubscript{20.1} film, whose highest measured resonance was \(7.4~\mathrm{GHz}\), the attenuation grows with frequency to a power-absorption ratio of \(0.076\) at that frequency (Fig.~\ref{fig4}a), about 38 times the \(\approx 2\times10^{-3}\) reported for a Permalloy film of the same \(400~\mu\mathrm{m}\) length at \(1.6~\mathrm{GHz}\)~\cite{Kurimune2020}. Frequency enhancement thus lifts weakly magnetoelastic Permalloy to the attenuation of the strongly magnetoelastic Ni film in that study, compensating for a small magnetoelastic coefficient.  

The resulting \(P(f)\), fitted by \(P(f) = \zeta f^{3}\) (Fig.~\ref{fig4}b--g), yields a \(\zeta\) that increases systematically towards Fe-rich alloys (Fig.~\ref{fig4}h). Evaluating Eq.~(\ref{eq:main2}) at \(\phi = 0\) with the ellipticity of Eq.~(\ref{eq:main4}) and the rotation-based \(\delta\) gives \(P(\phi = 0) \approx \zeta f^{3}\) with \(\zeta \propto M_{\text{S}}/\alpha\) (Supplementary Eq.~(S57)): the \(b^{2}\) in \(C_{0}\) cancels the \(1/b^{2}\) from \((1/\delta)^{2}\), leaving \(\zeta\) free of explicit \(b\). Here \(\alpha\) is the Gilbert damping, extracted for each composition from the linewidth (Supplementary Fig.~S3), which fixes \(\kappa_{\text{SW}} \approx \alpha|\gamma|\mu_{0}M_{\text{S}}/2\). The measured \(\zeta(x)\) and \(M_{\text{S}}/\alpha\) agree qualitatively (Fig.~\ref{fig4}h).

According to Eq.~(\ref{eq:main2}), \(\kappa_{\text{SW}}\) is roughly independent of \(f\) for \(2\pi f \ll |\gamma|\mu_{0}M_{\text{S}}\), while the same inequality makes the spin-wave precession strongly elliptic [Eq.~(\ref{eq:main4})], contributing a factor \(f^{-1}\). Substituting this into the MR \(\delta\) of Eq.~(\ref{eq:main3}) gives \(\delta \approx -2|\gamma|b\eta/(\pi f M_{\text{S}})\) and hence \(P(\phi = 0) \propto f^{3}\), consistent with experiment. The Barnett \(\delta\) of Eq.~(\ref{eq:main5}) shares the same \(f^{-1}\) dependence, giving the same \(f^{3}\) law and differing only by a factor of two; given the uncertainties in \(b\), \(M_{\text{S}}\), and \(\eta\), the data are consistent with either mechanism and cannot discriminate between them. This corrects the previous study~\cite{Xu2020}, which held the Barnett amplitude to be much smaller than the MR field: that argument overlooked the ellipticity. Because the MR and Barnett fields are polarized out of and in plane, respectively, their effects on the strongly oblate spin-wave mode are suppressed and enhanced, cancelling the amplitude difference.

At higher frequencies, where the condition \(2\pi f \ll |\gamma|\mu_{0}M_{\text{S}}\) is no longer valid, the ellipticity approaches unity, \(m_{\text{ip}}/m_{\text{oop}} \to 1\), and the spin-wave precession becomes circular. The MR \(\delta\) depends on frequency only through the ellipticity and therefore stops decreasing, whereas the Barnett \(\delta\) keeps its explicit \(f^{-1}\) from the additional time derivative of the spin-vorticity coupling. Through Eq.~(\ref{eq:main2}) the Barnett-induced \(P\) then grows two powers of \(f\) faster than the MR-induced \(P\), so their frequency dependences separate and the two mechanisms become distinguishable.
Within the \(\lesssim 9\)~GHz range accessed here, however, \(2\pi f\) remains well below \(|\gamma|\mu_{0}M_{\text{S}}\), so this separation remains unresolved.

\section*{Conclusions}
\hspace{15pt}Our measurements rule out the shear-strain mechanism as the origin of the magnetoacoustic nonreciprocity in the present experimental setup, since in an isotropic polycrystalline magnet it shares the single coefficient \(b\) with the longitudinal-strain drive and therefore could not produce the observed sign reversal. The nonreciprocity is consistent with a rotation-based mechanism instead. However, the present data cannot differentiate between the two rotation-based couplings, namely the Barnett field mechanism and the MR mechanism, as both agree with the observed cubic frequency scaling; resolving the two will require higher-frequency measurements or a magnetically isotropic film~\cite{Kurebayashi2026}, which suppresses the MR drive and leaves the Barnett effect alone. A magnetic field inductively generated by the SAW's piezoelectric field~\cite{Kawada2024EM} shares the same symmetry as the Barnett mechanism, and likely a comparable frequency dependence; in a different heterostructure this contribution was found to be comparable to the Barnett field, though its magnitude depends sensitively on the conductivity and geometry of the metallic stack, and a quantitative estimate for the present Ti/\NiFe{}/Ti film is left for future work. Tuning the alloy composition provides a materials-level handle on both the sign and the magnitude of magnetoacoustic nonreciprocity, broadening the design space for nonreciprocal magnon--phonon devices such as acoustic isolators~\cite{Shao2020}, on-chip circulators, and acoustically assisted memories~\cite{Camara2019}. Furthermore, the sub-micrometre-wavelength SAW devices boost the resonance frequency, which sustains an appreciable signal driven by the rotation-based coupling in Ni\textsubscript{79.9}Fe\textsubscript{20.1}, where \(b\) nearly vanishes; this allows us to still distinguish the sign of \(b\) via the nonreciprocity, establishing our method as a probe of the sign of \(b\) where strain-based metrology does not work. Residual magnetostriction makes the magnetic anisotropy sensitive to mechanical stress arising during fabrication, in particular heat treatment, generates magnetoelastic acoustic noise and vibrations~\cite{KlokholmAboaf1981,Freitas2007}, and produces additional stress upon magnetization switching. The sign readout demonstrated here provides the direction of the compositional correction that suppresses these effects and improves MRAM reliability~\cite{Engel2005,Sankaran2018}.

\section*{Methods}

\subsection*{Sample preparation}

\hspace{15pt}Figure~\ref{fig1}a illustrates the hybrid device architecture, which consists of three main components: a piezoelectric substrate (128\textdegree{} Y-X cut black LiNbO\textsubscript{3}, selected for its superior thermal stability compared to standard LiNbO\textsubscript{3}), a pair of interdigital transducers (IDTs) for the excitation and detection of surface acoustic waves (SAWs), and a rectangular trilayer heterostructure composed of Ti (3 nm) / \NiFe{} (10 nm) / Ti (3 nm) with a lateral size of \(80 \times 400~\mu\mathrm{m}^2\). The IDTs are designed to launch an acoustic beam with an aperture of \(80~\mu\mathrm{m}\), matched to the film width so that the propagating SAW is fully intercepted by the magnetic layer. The IDT fingers are made of Al (30 nm), while the contact pads are composed of Ti (3 nm) and Au (200 nm).

Device fabrication was carried out in an ISO Class 5 cleanroom following a seven-step process. First, the LiNbO\textsubscript{3} substrate was cleaned in an ultrasonic bath using acetone (5 min) and isopropanol (2 min). Second, the IDTs were patterned via electron beam lithography (Elionix7700) using PMMA A4 and E-spacer, with a designed finger periodicity of 630 nm. Third, Al was deposited by electron beam evaporation at 1 \(\times\) 10\textsuperscript{\(-\)4} Pa, followed by lift-off. Fourth, the contact pad regions were patterned by UV lithography using HMDS and AZ1500. Fifth, Ti/Au was deposited at 2 \(\times\) 10\textsuperscript{\(-\)4} Pa via electron beam evaporation, followed by lift-off. Sixth, the heterostructure region was defined using UV lithography. Finally, the Ti/\NiFe{}/Ti trilayer was deposited under a base pressure of 2 \(\times\) 10\textsuperscript{\(-\)4} Pa and patterned by lift-off.

A series of \NiFe{} films was prepared with Ni concentrations \emph{x} of 86.6, 83.5, 80.7, 79.9, 77.2, 76.1, 73.6, 73.4, and 66.3~at.\%. The metal sources (4N purity) were purchased from Furuuchi Chemical. For compositional analysis, reference films with the same Ti/\NiFe{}/Ti trilayer structure (10 mm \(\times\) 10 mm) were simultaneously deposited on SiO\textsubscript{2}/Si substrates. Film composition was determined by X-ray fluorescence (XRF). For the frequency-dependent measurements, IDTs of different periodicity were designed to launch SAWs with wavelengths down to \(\sim\)400 nm, covering acoustic resonance frequencies of \(\sim\)2--9 GHz; all devices of a given composition were patterned together on a single chip in one processing run to minimize the variation of film composition and microstructure across the series.

\subsection*{Measurement setup}

\hspace{15pt}The experimental setup is illustrated in Fig.~\ref{fig1}a. The device is mounted on a motorized stage capable of applying a uniform in-plane magnetic field with adjustable orientation and tunable strength ranging from \(-\)200 mT to 200 mT. The two terminals of the device are connected to port 1 and port 2 of a vector network analyser (VNA), which measures the radio-frequency transmission in both directions, corresponding to the scattering parameters $S_{21}$ and $S_{12}$. These parameters quantify the transmission ratio when the excitation is applied at port 1 or port 2, respectively. The Ni\textsubscript{79.9}Fe\textsubscript{20.1} device was measured at EPFL using a Keysight PNA N5222A network analyser, and the remaining devices were measured at RIKEN using a Keysight PNA N5225B network analyser. The VNA output power was 10~dBm. Surface acoustic waves (SAWs) are launched from the input port, propagate through the ferromagnetic heterostructure, and are converted back into electrical signals at the output port via the piezoelectric effect. The $S_{21}$ and $S_{12}$ signals correspond to SAWs propagating in the $+k$ and $-k$ directions, respectively; owing to the symmetric device layout, interchanging the input and output IDTs in this way probes the two propagation directions under otherwise identical conditions. The raw scattering parameters are recorded as linear magnitude and phase. The magnetoacoustic attenuation ratios are defined as \(A_{\text{+}k}(H) = \left\{ |S_{21}^{0}|^{2} - |S_{21}(H)|^{2} \right\}/|S_{21}^{0}|^{2}\) and \(A_{\text{-}k}(H) = \left\{ |S_{12}^{0}|^{2} - |S_{12}(H)|^{2} \right\}/|S_{12}^{0}|^{2}\), where \(|S_{21}^{0}|\) and \(|S_{12}^{0}|\) are reference values measured at the maximum magnetic field of the sweep, at which the magnetoacoustic attenuation is negligible. To suppress spurious contributions from direct electromagnetic crosstalk, a time-gating technique is employed: a radio-frequency pulse is applied to the input, and the output signal is recorded in the time domain. The early-arriving electromagnetic component is removed, and the remaining acoustic signal is Fourier-transformed into the frequency domain. The acoustic resonance condition is \(f_{0} = v/\lambda_{\text{SAW}}\), where \(v\) is the SAW velocity and \(\lambda_{\text{SAW}}/2\) equals the centre-to-centre spacing between the signal and ground electrodes. In the experiment, the actual value of \(f_{0}\) is determined by sweeping the excitation frequency and identifying the maximum in transmission, which in our devices occurs around 6.2 GHz. Field-swept attenuation spectra were recorded with step sizes of 0.5--1 mT depending on the device. For the angular measurements, spectra were recorded at \(5^{\circ}\) steps spanning \(180^{\circ}\) to \(360^{\circ}\); the peak absorption \(P_{\text{+}k}\) at positive field (\(\phi \in [180^{\circ}, 360^{\circ}]\)) was obtained by Lorentzian fitting [Eq.~(\ref{eq:main1})], and the \(P_{\text{-}k}\) data measured between \(180^{\circ}\) and \(355^{\circ}\) were mapped to \(0^{\circ}\)--\(175^{\circ}\) assuming a symmetric setup. All measurements were performed at room temperature.

\section*{Data availability}

The data presented in the main text and the supplementary information are available from the corresponding authors upon request.

\section*{Code availability}

The code used to fit the magnetoacoustic attenuation spectra and to extract the parameters presented in the main text and the supplementary information is available from the corresponding authors upon request.

\bibliography{referencesV3}

\section*{Acknowledgments}

The authors express their gratitude to Olivier Klein, Takeshi Seki, Jorge Puebla, James Jun He, and Gerrit Bauer for their insightful discussions. This work was supported by the JST ASPIRE Program (Grant No. JPMJAP2410) and by the Grant-in-Aid for Transformative Research Areas (A) ``Physics of Chimera Quasiparticles'' (Grant No. 24H02233). M.X. acknowledges support from JSPS through the ``Research Program for Young Scientists'' (No. 19J21720) and the RIKEN IPA Program. S.M. acknowledges support from JSPS KAKENHI (Grant No. 24K00576 from MEXT, Japan). K.Y. acknowledges support from JSPS KAKENHI (Grant Nos. 21K13886, 25H00837, and 26K07006), JST PRESTO (Grant No. JPMJPR20LB), and the JSPS Bilateral Program (Grant No. JPJSBP120245708).

The use of the facilities of the Emergent Matter Science Research Support Team in RIKEN is gratefully acknowledged.

\section*{Author contributions}

M.X., K.K., and Y.O. conceived the project. M.X. and K.Y. wrote the main manuscript and the Supplementary Information. M.X. fabricated the samples and performed the acoustic absorption measurements. K.Y. formulated the theoretical model. M.X. and K.Y. analysed the data. T.K., K.A., and D.H. characterized the composition. Z.Z. and M.X. characterized the magnetic properties of the film. M.X., K.Y., K.K., Z.Z., K.A., T.K., D.H., L.L., D.G., S.M., and Y.O. discussed the results and contributed to the manuscript. K.K., S.M., and Y.O. supervised the project. All authors discussed the results and commented on the manuscript.

\section*{Competing interests}

The authors declare that they have no competing interests.

\section*{Figures}

\end{document}